\pdfoutput=1
\documentclass[11pt, letterpaper]{article}
\usepackage[top=1in, bottom=1in, left=1in, right=1in]{geometry}

\usepackage{amsmath,amssymb,amsfonts,mathrsfs,mathtools}
\usepackage{graphicx}
\usepackage[round,numbers,sort&compress]{natbib}

\usepackage[T1]{fontenc}
\usepackage{lmodern}
\usepackage[sc]{mathpazo}
\usepackage{textcomp}

\usepackage[labelfont={bf}, margin=1cm]{caption}
\usepackage{units}

\usepackage{authblk}

\newcommand{\splitcell}[2][c]{%
  \begin{tabular}[#1]{@{}l@{}}#2\end{tabular}}

\begin{document}
\title{A Coupled Stochastic Model Explains Differences\\ in Circadian Behavior of \textit{Cry1} and \textit{Cry2} Knockouts}
\author[1]{John~H.~Abel}
\author[2]{Lukas A. Widmer}
\author[1]{Peter~C.~St.~John}
\author[2]{J{\"o}rg Stelling}
\author[1,*]{Francis~J.~Doyle~III}
\affil[1]{Department of Chemical Engineering, University of California Santa Barbara, Santa~Barbara, California 93106-5080}
\affil[2]{Department of Biosystems Science and Engineering, 
ETH Z{\"u}rich, and Swiss~Institute~of~Bioinformatics, 4058 Basel,  Switzerland}
\affil[*]{Email: \texttt{doyle@engineering.ucsb.edu}}
\date{\today}

\maketitle

\begin{center}
Classification:\\ Quantitative Biology | Molecular Networks \\[1ex]
\end{center}

\pagebreak

\let\thefootnote\relax\footnote{This work was supported in part by the National Institutes of Health under grant 1R01GM096873-01, and the Institute for Collaborative	Biotechnologies under grant W911NF-09-0001 from the U.S. Army Research Office. The content of the information does not necessarily reflect the position of the policy of the Government, and no official endorsement should be inferred.}

\section{Introduction}
Circadian
rhythmicity in gene expression affects a wide variety of biological processes at multiple scales, including sleep-wake cycles, feeding-fasting behavior, cell cycles, and metabolism \cite{Bass2010, Bieler2014, Green2008a}. 
A population of approximately 20,000 oscillators in the suprachiasmatic nucleus of the brain serves as the mammalian master clock, and synchronizes circadian oscillations in peripheral tissues \cite{Lamia2008a}.
Cellular oscillators in the SCN display stochastic variation in rhythmicity
and period length, due to the low molecular counts of core clock proteins.
Stochastic noise has previously been implicated in a variety of phenomena
within the SCN \cite{Webb2009, Ko2010}, and is thought to
play an important role in SCN behavior \cite{Forger2005}.
SCN cells resist noise to establish robust system-wide rhythmicity through coupling via the neurotransmitter VIP, which promotes transcription of the circadian gene \textit{Period} (\textit{Per}) \cite{Aton2005, Welsh2010}.

The mammalian cell-autonomous oscillator is driven by a central transcription-translation feedback loop, in which the protein products of E box-activated genes \textit{Per} and \textit{Cryptochrome} (\textit{Cry}) form heterodimers to repress their own transcription, as shown in Fig. \ref{fig:model}.
As these transcription factors are degraded, transcription begins again, repeating the cycle with near-24 hour periodicity.  
While both CRY1 and CRY2 isomers are known E box repressors,
\textit{Cry1} and \textit{Cry2} knockouts display significant differences
in rhythmicity and period length \cite{Liu2007, VanderHorst1999}.  Prior
work has shown that single dissociated SCN neurons display persistent
circadian rhythmicity in wild-type and \textit{Cry2} knockout,
while dissociated \textit{Cry1} knockout cells are largely arrhythmic.
However, rhythmicity may be restored in \textit{Cry1} knockout populations
through intercellular coupling \cite{Liu2007, Evans2012}. 
This difference in knockout behaviors has led to speculation that CRY1 and CRY2 play distinct, non-redundant roles in the circadian oscillator or differ in strength of repression \cite{Ukai-Tadenuma2011,McCarthy2009, Khan2012}. 

Here, we address the roles of \textit{Cry1} and \textit{Cry2} through a
coupled stochastic model of the core circadian oscillator, explicitly
accounting for intrinsic noise and coupling dynamics through the VIP and
C-response element binding protein (CREB) pathway. 
We demonstrate that it is not necessary for CRY1 and CRY2 to perform different functions in the oscillator; rather, knockout characteristics of SCN neurons are well-explained by the absolute abundance of PER, CRY1, and CRY2 proteins, in conjunction with intrinsic noise. 
Specifically, the higher expression of \textit{Cry1} \cite{Lee2001a} causes \textit{Cry1} knockouts to lack sufficient CRY to rhythmically repress transcription.
When coupled, the positive-feedback VIP loop forces cells to cross a bifurcation and become limit cycle oscillators, as positive feedback loops under certain conditions promote oscillatory behavior \cite{Ananthasubramaniam2014a}.
The results reconcile single uncoupled cell and network behavior, and demonstrate the importance of absolute abundances and stochastic noise in understanding the circadian oscillator.

\section{Model Structure and Parameter Identification}
\begin{figure}[tbp]
	\begin{center}
	\includegraphics[width=3.5in]{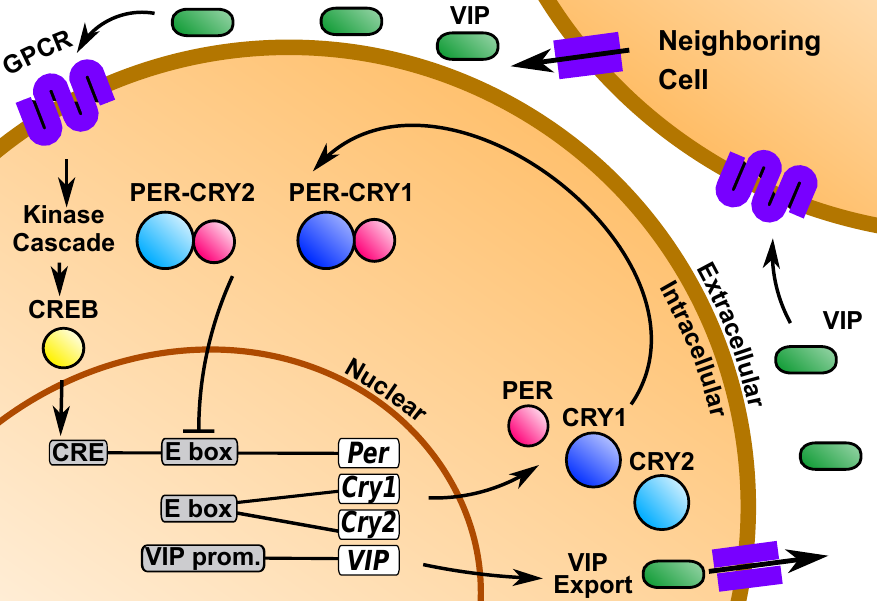}
	\end{center}
	\caption{Visualization of model components. The model explicitly
		includes four mRNA species (\textit{Per, Cry1, Cry2, VIP}), four cytosolic proteins (PER, CRY1, CRY2, VIP), and three nuclear transcription regulators (PER-CRY1, PER-CRY2, and CREB). The central negative feedback loop (via PER-CRY) is modulated by the external positive feedback loop (via VIP) which mediates cell to cell communication. VIP excreted from nearby cells is received by G protein-coupled receptor (GPCR), which couples cells in proximity.}
		\label{fig:model}
\end{figure}

A schematic representation of the states captured in this model is shown in 
Fig. \ref{fig:model}. We focus on accurately capturing the network
dynamics of the central \textit{Per-Cry} transcription-translation feedback
loop.
For intercellular coupling, we include the VIP-CREB
pathway, as VIP has been shown to play an essential role in driving
synchronization within the SCN through induction of the core clock gene
\textit{Per} \cite{Aton2005, Welsh2010}.
Here, we assume that VIP production is modulated by the PER-CRY complex, as it is synthesized in phase with \textit{Cry} mRNAs \cite{Lee2001a}, and is regulated by the clock.
In \cite{Ananthasubramaniam2014}, the phase of VIP release was found to be important for driving synchronization.
The model agrees with both this theory and experiment in that VIP is released in-phase with \textit{Per} expression \cite{Morin1991}.

The model explicitly includes the dynamics of the PER-CRY1-CRY2 feedback loop and VIP signalling, resulting in an 11-state model with 34 kinetic parameters.
The differential equations and the corresponding stochastic propensity functions for each state were formulated using Michaelis-Menten-type kinetics for repression and enzyme-mediated degradation, and mass-action kinetics for translation and dimerization reactions.
For \textit{Per}, an additional CREB-dependent activation term was added to the Michaelis-Menten repression. 
Differences between \textit{Cry1} and \textit{Cry2} expression levels are captured through differing transcription, translation, and degradation rates. 
Despite some evidence suggesting a difference in repressive ability between CRY isoforms \cite{GriffinJr.1999}, the model did not require explicitly differentiating between repressive potency to accurately capture experimental phenotypes.
While repressive potency could compensate for protein concentration differences in simplified deterministic models, molecular noise is intrinsically tied to absolute molecular numbers.
Therefore in biochemical systems where stochastic effects are relevant,
repressive potency is inherently different from relative abundance.

We optimized model parameters using an evolutionary algorithm.
The optimization was performed with respect to experimental mRNA and protein concentrations and amplitudes, RNAi sensitivities, and phase relationships \cite{Lee2001a,Lee2011, VanderHorst1999, Obrietan1999, Tischkau2003, Nielsen2002, Zhang2009}.
Though quantitative data exists only for liver and fibroblast cells, we assume relative mRNA and protein concentrations are conserved between cell types \cite{Lee2001a,Lee2011}.
Parameters were estimated from the deterministic model.
Since a majority of quantitative biological data exist at population level, we optimized the model in the coupled state with the VIP pathway included.
To model uncoupled cells, all VIP concentrations were set to zero, which reflects the low concentration of VIP and lack of functional synapses in cultures of mechanically-dissociated SCN neurons.
Cells within the coupled population were connected to horizontal and vertical neighbors on a 15-by-15 cell grid, with periodic boundaries.
For conversion to a stochastic model, the volume parameter $\Omega$ was fit through the desynchronization rate of decoupled oscillators to correctly capture intrinsic stochastic noise \cite{Rougemont2007, St.John2014}.

The model equations for each state were written and solved using the CasADi computer algebra package \cite{Andersson2012} and the SUNDIALS ODE solvers suite \cite{Hindmarsh2005}.
The model was simulated stochastically using the Gillespie algorithm, as implemented in StochKit2 \cite{Sanft2011}.
A detailed description of model equations, parameters, and fitting is available \cite{Abel2014}.

\section{Results}
\begin{figure}[tb]
	\begin{center}
	\includegraphics[width=3.5in]{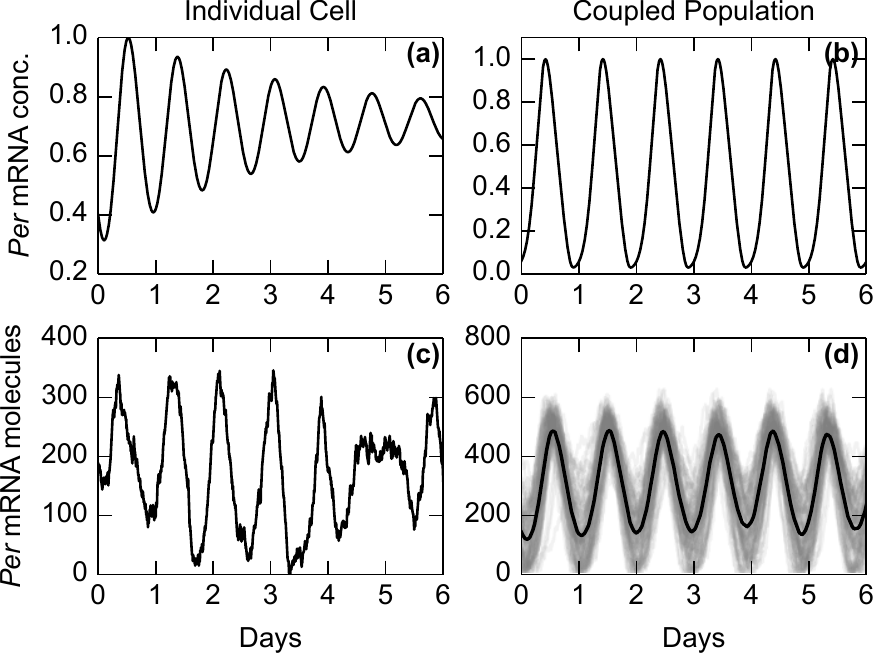}
\end{center}
	\caption{Deterministic ((a)-(b)) and stochastic ((c)-(d)) simulation of the single uncoupled cell and
		network-level model. Deterministic trajectories represent normalized \textit{Per} mRNA concentrations, stochastic trajectories show \textit{Per} mRNA molecule count. (a) The deterministic single uncoupled cell model shows damped oscillations.
		(b) When coupling (VIP) is included, the model crosses a bifurcation to a deterministic limit-cycle oscillator. 
		(c) Stochastic simulation of the single uncoupled cell model demonstrates no decay in noise-driven oscillation amplitude. 
		(d) Synchronization is maintained at a mean population level ($n=225$ cells, black) despite noisy individual cells (gray traces).}
\label{fig:wt}
\end{figure}

\begin{figure}[tb]
	\begin{center}
	\includegraphics[width=3.5in]{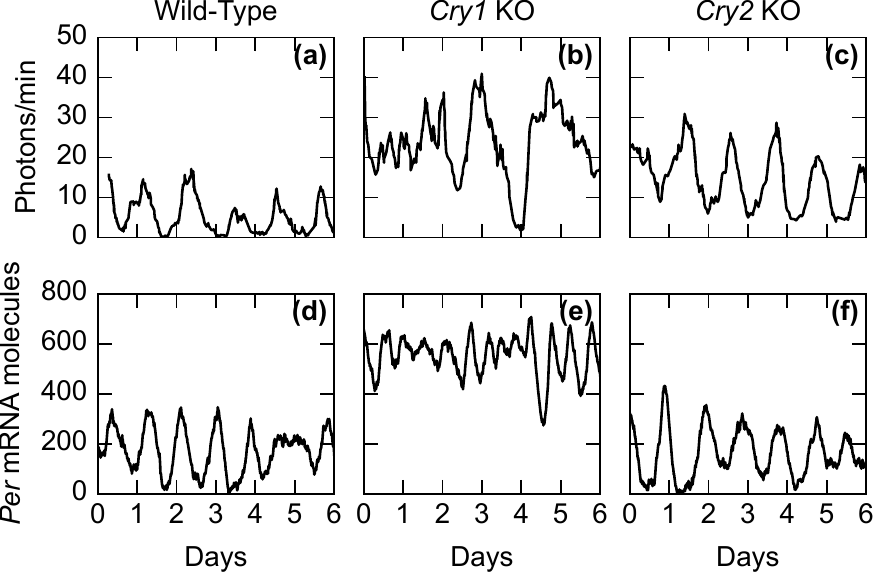}
	\end{center}
	\caption{
The stochastic mathematical model captures individual neuron behavior. 
(a)-(c) Experimental \textit{Per2-luc} traces, from \cite{Liu2007}. 
(d)-(f) \textit{Per} mRNA stochastic simulation trajectories from the model. 
}
	\label{fig:sc}
\end{figure}

\begin{figure}[tb]
	\begin{center}
	\includegraphics[width=3.5in]{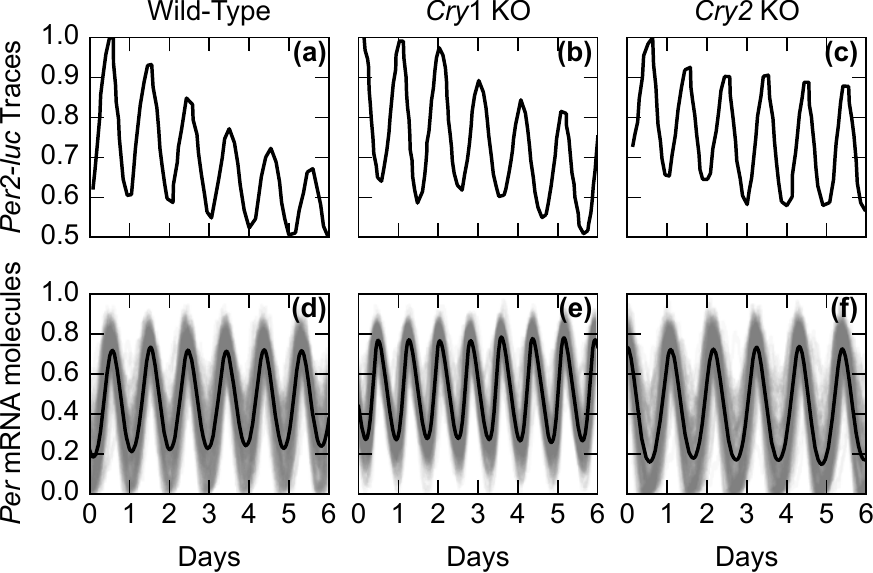}
	\end{center}
	\caption{
    Rhythmicity and synchronization were achieved for coupled SCN phenotypes in both experimental SCN explants \cite{Liu2007} ((a)-(c)) and \textit{in silico} coupled simulations ((d)-(f)). 
	Individual cell traces for simulation data are shown in gray. 
  Each trajectory shown is normalized, as the number of cells differs between experimental SCN explants. 
  A downward trend is observed in biological SCN explant rhythms likely due to cell death.
}
\label{fig:pop}
\end{figure}

In the single uncoupled cell case, the deterministic model shows damped oscillations (Fig. \ref{fig:wt}(a)). 
When simulated stochastically (Fig.~\ref{fig:wt}(c)), single dissociated cells display sustained non-damping oscillatory behavior.
This indicates that single uncoupled cells in this model are noise-driven oscillators, i.e. they are damped in the absence of noise; however, molecular noise in the biochemical reactions causes sustained oscillation \cite{Thomas2013}.
As shown in \cite{Westermark2009}, single dissociated neurons are well approximated as damped oscillators driven by molecular noise, and cannot be distinguished from noisy limit-cycle oscillators.
With the VIP coupling pathway included, the system crosses a bifurcation to a deterministic limit cycle oscillator (Fig.~\ref{fig:wt}(b)).
This bifurcation is caused by inclusion of the VIP positive feedback on \textit{Per} expression.
For uncoupled cells, this loop is knocked out.
That is, when uncoupled, each individual cell is a noise-driven
oscillator. When coupled, each individual cell becomes a limit
cycle oscillator.
When simulated stochastically in Fig.~\ref{fig:wt}(d), this coupling is sufficient to drive phase synchrony in the mean (black), despite noise in the single-cell trajectories (gray traces).

The central clock in the SCN has shown a strong robustness to genetic perturbations, and the ability to maintain synchronized rhythmicity even with knockouts to core clock genes \cite{Liu2007, Ko2010}. 
At a single dissociated cell level the clock is less robust to perturbation, with dissociated \textit{Cry1} knockouts displaying less regular rhythms. 
First we demonstrate that the model captures the phenotypes of \textit{Cry1} and \textit{Cry2} knockouts at both the level of dissociated cells (Fig.~\ref{fig:sc}) and coupled networks (Fig.~\ref{fig:pop}).
If both CRY isoforms act as E-box repressors, knockout of either \textit{Cry} is expected to increase transcription of \textit{Per}, with the response being more severe for the \textit{Cry1} knockout, as CRY1 is more abundant than CRY2. 
This is indeed seen in both \textit{Per2-luc} measurements \cite{Liu2007} and \textit{in silico Per} mRNA trajectories.
When coupling is restored as in full SCN explants, rhythmicity and synchronization is restored for \textit{Cry1} knockouts in both experiment and simulation, as shown in Fig.~\ref{fig:pop}. 
The model also correctly captures the directions of population-level period sensitivity to \textit{Cry} knockouts.
Model \textit{Cry1} knockouts displayed a shortened period (82\% of wild-type length), and model \textit{Cry2} knockouts displayed a lengthened period (114\% of wild type), consistent with period shortening and lengthening in experiment (\textit{Cry1} knockout 90-95\%, \textit{Cry2} knockout 105-110\%) \cite{VanderHorst1999,Lee2001a,Liu2007}.

To quantify neuron behavior, we use a rhythmicity index as defined in \cite{Leise2011}: the fraction of energy contained within the frequency band corresponding to circadian oscillations, as calculated via a discrete wavelet transform (DWT). 
The discrete wavelet transform is performed by applying high and low-pass filters consecutively to resolve a time series at different scales. 
The energy $\|D_j\|$ for each DWT level $j$ is
\begin{equation}
	\|D_j\|^2 = \sum_{k=1}^K W_{j,k}^2,
	\label{eqn:power}
\end{equation}
where $W_{j,k}$ is the $k^{th}$ wavelet coefficient at DWT level $j$.
Thus, the rhythmicity index is
\begin{equation} RI =
	\frac{\|D_{c}\|^2}{\sum\limits_{j=1}^J \|D_j\|^2},
	\label{eqn:ri}
\end{equation} 
where $J$ is the number of levels of the discrete wavelet transform, and $D_c$ is the energy of the wavelet level containing the circadian frequency.
By Parseval's theorem, this is equivalent to the mean of the squared Fourier coefficients within each frequency band \cite{Leise2011}.
We define the circadian frequency band $D_{c}$ to contain periods between $\frac{2}{3}$ and $\frac{4}{3}$ of mean wild-type period, as it will capture oscillations with a near-circadian frequency.

\begin{figure}[tb]
	\begin{center}
	\includegraphics[width=3.5in]{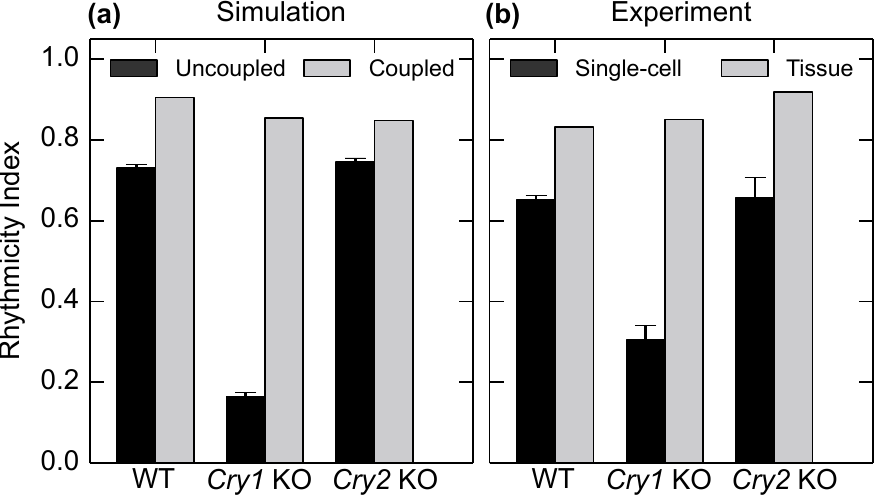}
	\end{center}
	\caption{Model correctly predicts noise characteristics of single uncoupled cells (black) and the coupled population (gray). Error bars represent variance.
		(a) Stochastic simulation. For each sample, $n=225$.  
		(b) Experimental results from \cite{Liu2007}. Single dissociated cell mean and variance were taken from ten cells of each phenotype. SCN tissue rhythmicity was taken from tissue explants of $O(10^3)$ cells. Overall knockout behavior corresponds closely with model prediction.}
	\label{fig:dwt}
\end{figure}

When this metric is applied (Fig. \ref{fig:dwt}), experimental and simulated results show good agreement, and demonstrate the strong increase in circadian rhythmicity in coupled \textit{Cry1} knockouts. 
Since \textit{Cry1} and \textit{Cry2} play the same role in the model, the result indicates that the relative abundance of \textit{Cry} isoforms in conjunction with intrinsic molecular noise is sufficient to capture the differing effects of \textit{Cry1} and \textit{Cry2} knockouts. 

\section{Conclusion}
In this work, we applied a new coupled stochastic model of the circadian oscillator to a lingering question regarding the roles of \textit{Cry} isoforms.
We demonstrate that relative abundance is sufficient to explain \textit{Cry} knockout behaviors, and that a parallel role for \textit{Cry1} and \textit{Cry2} in the circadian gene repression is consistent with experimental data.
These results suggest that the inclusion of VIP coupling causes individual cells to cross a Hopf bifurcation.
This further supports the possibility that single, decoupled cells may not exhibit deterministic limit cycles, though single-cell limit cycle oscillators remain a widely-used convention.
While deterministic models are sufficient to capture population-level limit cycle behavior, they are insufficient to describe the noise-induced oscillations in single dissociated cells.
Finally, these results emphasize the importance of describing circadian rhythmicity at the single-cell level with a continuous metric, rather than the traditional binary classification of ``rhythmic'' or ``arrhythmic.''
Though the chosen rhythmicity metric is not new, it adds detail to the understanding of cellular circadian behavior, and further validates the dynamics captured by this new model.

\bibliographystyle{myIEEEtran.bst}
\bibliography{condensed_library.bib}


\section{Appendix: Model Description}
\subsection*{Deterministic Model}
The deterministic model is composed of eleven ODEs.
Each reaction in the model obeys either mass-action or Michaelis-Menten kinetics, and the system behaves as a limit cycle oscillator without requiring higher-order Hill kinetics.
This reflects \cite{Ananthasubramaniam2014a}, in that a positive feedback loop (VIP-CREB promotion of \textit{Per} transcription) ameliorates the need for physically unrealistic high order Hill terms.
Avoiding Hill terms decreases error in the stochastic simulation, as the linear noise approximation (LNA) variance of the stochastic Michaelis-Menten equation is known to be bounded to $[1,\frac{4}{3}]$ of the true LNA variance at steady state \cite{Widmer2013}.

The period of the deterministic model was scaled to 23.7 hours, however, the amplitudes were not scaled. 
The units of the state variables are arbitrary relative concentrations, and the units of parameters are provided.
Tables \ref{tab:eqn} and \ref{tab:par} contain the equations and parameters, respectively, for the ODE model of circadian cells.

\subsection*{Model Assumptions}
The following assumptions were made in the development of this model:
\begin{itemize}
	\item CLOCK and BMAL1 are considered approximately constant, through the $v_{2pr}$ parameter. This parameter captures the repression of E-box genes, which is achieved through binding of PER-CRYs to CLOCK-BMAL1.
	\item Degradation within the cell is assumed to be enzyme-mediated with Michaelis-Menten kinetics. Extracellular VIP is assumed to degrade with mass-action kinetics.
	\item No Hill term of order higher than unity is used, as the external positive feedback arm causes sufficient nonlinearity to create a limit cycle.
	\item Because nuclear CRY1 and CRY2 are believed to share a degradation pathway, shared degradation kinetics are derived under the pseudo-steady state hypothesis, as in \cite{Hirota2012a}.
	\item PER is assumed to be required for nuclear uptake of CRY. We additionally assume that equilibrium between PER-CRY complexes and free CRY proteins in the nucleus is fast, and do not explicitly consider multiple CRY nuclear states.
\end{itemize}

\subsection*{Conversion to Coupled Stochastic Form}

\begin{figure}
\begin{center}
        \includegraphics{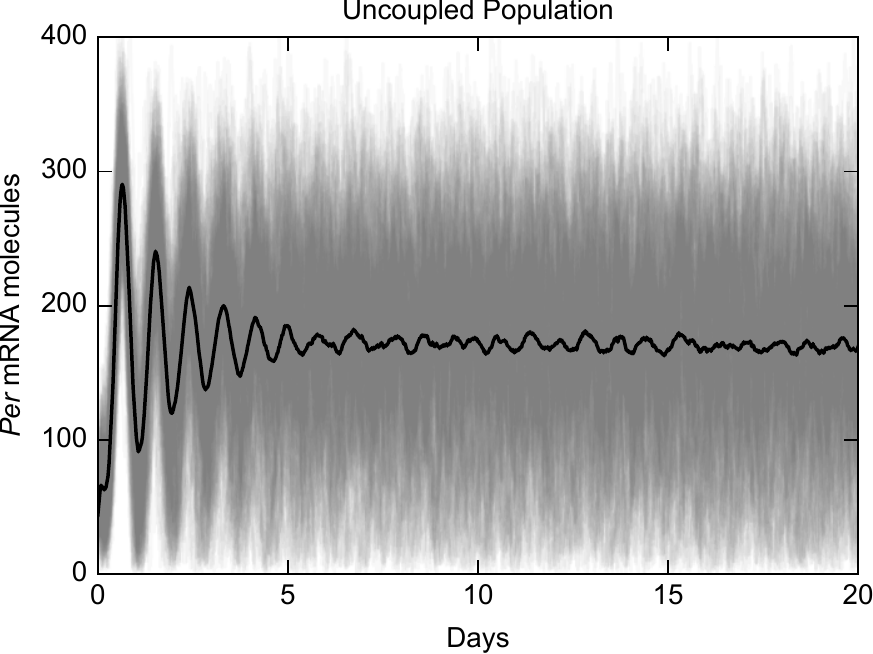}
    \end{center}
    \caption{A population of 225 uncoupled cells results in gradual desynchronization and eventually low-amplitude fluctuations caused by random noise.
    Individual cells show a wide variance in period length with no VIP coupling. This variance leads to gradual desynchrony.
Though the population average displays only low-level fluctuations (black line), individual cells continue to oscillate without damping (gray traces).}
\end{figure}

To create a stochastic form of the model, concentration was converted to population using a volume parameter $\Omega$.
This volume was determined by fitting the stochastic model to the desynchronization rate of a population of decoupled oscillators, to capture the correct amount of stochasticity in the population \cite{St.John2014}.
For the model, $\Omega = 400$.
Additionally, a diffusion rate between adjacent cells was defined, and an equation added to account for coupling between cells.
Networks for population simulations consisted of 225 cells. 
Coupled populations were placed on a 15-by-15 cell grid, and coupled to the two nearest neighbors horizontally and vertically. 
Boundary conditions were considered to be periodic.
VIP in the grid location of one cell ($VIP_1$) may move to a coupled cell ($VIP_2$) with propensity $k_t\times \mathbf{VIP_1} \Omega$, in units molecules/hour.
We used a $k_t$ value of 10.0 to reflect the fact that diffusion occurs rapidly in comparison to the slow reactions involved in cellular timekeeping.
However, $k_t$ values within an order of magnitude yield near-identical results.

\subsection*{Model Fitting}
The model was fit by use of an evolutionary algorithm in Python \cite{Fortin2012}.
Parameter values were generated randomly.
Criteria used in the fitness function constructed for this algorithm are shown in Table \ref{tab:fit}.
If the solution for a set of parameter values was not a limit-cycle oscillator, the fitness function returned a maximum value.

\bgroup
\def\arraystretch{1.3}
\begin{table}[htb]
\caption{Ordinary differential equations comprising the coupled circadian model.}
\label{tab:eqn}
\begin{center}
{
\begin{tabular}{l l l}\hline
	State Variable & Symbol & Model Equation\\
	\hline\\
	\textit{Per} mRNA & \textbf{p} & $\displaystyle\frac{d\mathbf{p}}{dt} = \frac{v_{1pp}\mathbf{CREB} + v_{2pr}}{K_{1p} + \mathbf{C1P}+ \mathbf{C2P}} - \frac{v_{3p}\mathbf{p}}{K_{2dp}+\mathbf{p}}$
	\\[0.6cm]
	\textit{Cry1} mRNA & \textbf{c1} & $\displaystyle\frac{d\mathbf{c1}}{dt} = \frac{v_{4c1r}}{K_{3c} + \mathbf{C1P}+ \mathbf{C2P}} - \frac{v_{5c1}\mathbf{c1}}{K_{4dc}+\mathbf{c1}}
$ 
	\\[0.6cm]
	\textit{Cry2} mRNA & \textbf{c2} & $\displaystyle\frac{d\mathbf{c2}}{dt} = \frac{v_{6c2r}}{K_{3c} + \mathbf{C1P}+ \mathbf{C2P}} - \frac{v_{7c2}\mathbf{c2}}{K_{4dc}+\mathbf{c2}}
$ 
	\\[0.6cm]
	\textit{VIP} mRNA & \textbf{vip} & $\displaystyle\frac{d\mathbf{vip}}{dt} = \frac{v_{8vr}}{K_{5v} + \mathbf{C1P}+ \mathbf{C2P}} - \frac{v_{9v}\mathbf{vip}}{K_{6dv}+\mathbf{vip}}
$ 
	\\[0.6cm]
	\textit{Per} Protein & \textbf{P} & \splitcell{
		$\displaystyle\frac{d\mathbf{P}}{dt} = k_{1p}\mathbf{p} - \frac{v_{10P}\mathbf{P}}{K_{8dP}+\mathbf{P}} - v_{11aCP}\mathbf{P}\times\mathbf{C1}- v_{11aCP}\mathbf{P}\times\mathbf{C2}$ \\[0.5cm] $\;\;\;\;\;\;\;\; + v_{12dCP}\mathbf{C1P} + v_{12dCP}\mathbf{C2P}
	$ }
	\\[1.1cm]
	\textit{Cry1} Protein & \textbf{C1} & $\displaystyle\frac{d\mathbf{C1}}{dt} = k_{2c}\mathbf{c1} - \frac{v_{13C1}\mathbf{C1}}{K_{9dC}+\mathbf{C1}} - v_{11aCP}\mathbf{P}\times\mathbf{C1} + v_{12dCP}\mathbf{C1P}
$ 
	\\[0.6cm]
	\textit{Cry2} Protein & \textbf{C2} & $\displaystyle\frac{d\mathbf{C2}}{dt} = k_{2c}\mathbf{c2} - \frac{v_{14C2}\mathbf{C2}}{K_{9dC}+\mathbf{C2}} - v_{11aCP}\mathbf{P}\times\mathbf{C2} + v_{12dCP}\mathbf{C2P}
$ 
	\\[0.6cm]
	\textit{VIP} Protein & \textbf{VIP} & $\displaystyle\frac{d\mathbf{VIP}}{dt} = k_{3v}\mathbf{vip} - v_{15V}\mathbf{VIP}
$ 
	\\[0.6cm]
	CRY1-PER Dimer & \textbf{C1P} & $\displaystyle\frac{d\mathbf{C1P}}{dt} = v_{11aCP}\mathbf{P}\times\mathbf{C1} - v_{12dCP}\mathbf{C1P} - \frac{v_{16C1P}\mathbf{C1P}}{K_{10dCn}+\mathbf{C1P} + \mathbf{C2P}}
$ 
	\\[0.6cm]
	CRY2-PER Dimer & \textbf{C2P} & $\displaystyle\frac{d\mathbf{C2P}}{dt} = v_{11aCP}\mathbf{P}\times\mathbf{C2} - v_{12dCP}\mathbf{C2P} - \frac{v_{17C2P}\mathbf{C2P}}{K_{10dCn}+\mathbf{C1P} + \mathbf{C2P}}
$ 
	\\[0.6cm]
	CREB Protein & \textbf{CREB} & $\displaystyle\frac{d\mathbf{CREB}}{dt} = \frac{v_{18V}\mathbf{VIP}}{K_{11V}+\mathbf{VIP}} - \frac{v_{19CR}\mathbf{CREB}}{K_{12dCR}+\mathbf{CREB}}
$ 
	
\end{tabular}
}
\end{center}
\end{table}

\begin{table}[htb]
\caption{Parameter descriptions for circadian ODE model.}
\label{tab:par}
\begin{center}
{
\begin{tabular}{l l l l}\hline
	Parameter & Description & Value & Units\\
	\hline
	$v_{1pp}$ & CREB-induced \textit{Per} mRNA promotion & 0.235 & [-]/hr
	\\
	$v_{2pr}$ & \textit{Per} mRNA transcription & 0.415 & [-]$^2$/hr
	\\
	$v_{3p}$ & \textit{Per} mRNA degradation & 0.478 & [-]/hr
	\\
	$v_{4c1r}$ & \textit{Cry1} mRNA transcription & 0.350 & [-]$^2$/hr
	\\
	$v_{5c1}$ & \textit{Cry1} mRNA degradation & 1.44 & [-]/hr
	\\
	$v_{6c2r}$ & \textit{Cry2} mRNA transcription & 0.124 & [-]/hr
	\\
	$v_{7c2}$ & \textit{Cry2} mRNA degradation & 2.28 & [-]/hr
	\\
	$v_{8vr}$ & \textit{VIP} mRNA transcription & 0.291 & [-]$^2$/hr
	\\
	$v_{9v}$ & \textit{VIP} mRNA degradation & 1.35 & [-]/hr
	\\
	$v_{10P}$ & \textit{Per} protein degradation & 13.0 & [-]/hr
	\\
	$v_{11aCP}$ & PER-CRY dimer formation & 0.493 & ([-]$\times$ hr)$^{-1}$
	\\
	$v_{12dcp}$ & PER-CRY dimer dissociation & 0.00380 & 1/hr
	\\
	$v_{13C1}$ & \textit{Cry1} protein degradation & 4.12 & [-]/hr
	\\
	$v_{14C2}$ & \textit{Cry2} protein degradation & 0.840 & [-]/hr
	\\
	$v_{15V}$ & \textit{VIP} protein degradation & 0.723 & 1/hr
	\\
	$v_{16C1P}$ & PER-CRY1 dimer degradation & 0.0306 & [-]/hr
	\\
	$v_{17C2P}$ & PER-CRY2 dimer degradation & 0.0862 & [-]/hr
	\\
	$v_{18V}$ & CREB activation by VIP receptors & 0.789 & [-]/hr
	\\
	$v_{19CR}$ & CREB deactivation & 1.27& [-]/hr
	\\
	$k_{1p}$ & PER translation & 7.51 & 1/hr	
	\\
	$k_{2c}$ & CRY translation & 0.572 & 1/hr
	\\
	$k_{3v}$ & VIP translation & 5.50 & 1/hr
	\\
	$K_{1p}$ & \textit{Per} transcription constant & 0.264 & [-]
	\\
	$K_{2dp}$ & \textit{Per} degradation constant & 0.00795 & [-]
	\\
	$K_{3c}$ & \textit{Cry} transcription constant & 0.156 & [-]
	\\
	$K_{4dc}$ & \textit{Cry} degradation constant & 1.94 & [-]
	\\
	$K_{5v}$ & \textit{VIP} transcription constant & 0.115 & [-]
	\\
	$K_{6dv}$ & \textit{VIP} degradation constant & 0.110 & [-]
	\\
	$K_{7dP}$ & \textit{Per} protein degradation constant & 0.0372 & [-]
	\\
	$K_{8dC}$ & \textit{Cry} protein degradation constant & 4.23 & [-]
	\\
	$K_{9dCn}$ & PER-CRY dimer degradation constant & 0.0455 & [-]
	\\
	$K_{10V}$ & CREB protein activation constant & 1.46 & [-]
	\\
	$K_{11CR}$ & CREB protein deactivation constant & 1.01 & [-]

\end{tabular}
}
\end{center}
\end{table}
\egroup

\bgroup
\def\arraystretch{1.3}
\begin{table}[htb]
\caption{Components of the model fitness function for optimizing a parameter set. Criteria 1-11, 14-18 from \cite{Lee2001a, Lee2011}; criteria 12, 13 from \cite{VanderHorst1999,Lee2001a,Liu2007}; criterion 20 from \cite{Morin1991}; and criteria 19, 21 from \cite{Obrietan1999a}.
}
\label{tab:fit}
\begin{center}
{
\begin{tabular}{l l l l l}\hline
Index & Description & Weight & Desired & Result\\
	\hline
\\
1&\textit{Per} mRNA Peak-trough ratio & $0.5$ & Large & Large\\
2&\textit{Cry1} mRNA Peak-trough ratio & $0.5$ & $2.16$ & $2.30$\\
3&\textit{Cry2} mRNA Peak-trough ratio & $0.5$ & $2.24$ & $2.20$\\
4&\textit{Per} protein Peak-trough ratio & $3$ & Large & Large\\
5&\textit{Cry1} protein Peak-trough ratio & $3$ & $3.247$ & $2.41$\\
6&\textit{Cry2} protein Peak-trough ratio & $3$ & $1.98$ & $1.60$\\
7&Fraction PER of total PER, CRY1, CRY2 & $3$ & $0.10$ & $0.06$\\
8&Fraction CRY1 of total PER, CRY1, CRY2 & $3$ & $0.56$ & $0.63$\\
9&Fraction CRY2 of total PER, CRY1, CRY2 & $3$ & $0.34$ & $0.31$\\
10&\textit{Cry1} siRNA sensitivity & $5$ & $<0$ & $<0$ \\
11&\textit{Cry2} siRNA sensitivity & $5$ & $>0$ & $>0$ \\
12&\textit{Cry1} knockout period & $12$ & $<0.95$ & $0.89$ \\
13&\textit{Cry2} knockout period & $12$ & $>1.15$ & $1.16$ \\
14&Fraction PER-CRY1 of total CRY1 & $1$ & $0.40$ & $0.22$ \\
15&Fraction PER-CRY2 of total CRY2 & $1$ & $0.35$ & $0.10$ \\ 
16&$t_{max}$ mRNAs - $t_{max}$ complexes & $3$ & $0.75$ & $0.81$ \\
17&$t_{max}$ cytocsolic protein - $t_{max}$ mRNAs & $3$ & $0.25$ & $0.01$ \\
18&$t_{max}$ PER-CRY - $t_{max}$ cytosolic protein & $3$ & $0$ & $0.18$ \\
19&$t_{max}$ CREB - $t_{max}$ mRNAs & $8$ & $>0.80$ & $0.93$ \\
20&\textit{VIP} protein peak-trough ratio & $5$ & $3.00$ & $3.48$ \\
21&CREB peak-trough ratio & $5$ & $3.00$ & $2.76$ \\
22&Deterministic coupled cells must synchronize & $20$ & True & True \\

\end{tabular}
}

\end{center}
\end{table}

\end{document}